\documentclass[fleqn,usenatbib]{mnras}
\usepackage[T1]{fontenc}
\usepackage{ae,aecompl}
\usepackage{graphicx}
\usepackage{amsmath}
\usepackage{amssymb}
\usepackage{natbib}
\usepackage{epstopdf}

\usepackage{float}

\usepackage{times}
\usepackage{amsmath}
\usepackage{amssymb}
\usepackage{amsfonts}
\usepackage{mathrsfs}
\usepackage{hyperref}

\DeclareRobustCommand{\VAN}[3]{#2}
\let\VANthebibliography\thebibliography
\def\thebibliography{\DeclareRobustCommand{\VAN}[3]{##3}\VANthebibliography}

\title [The age of IC~4665]{A revised age greater than 50~Myr for the young cluster IC~4665}
\author[R. D. Jeffries et al.]
  {R. D.~Jeffries$^1$, R.~J. Jackson$^1$ and A.~S.~Binks$^2$ \\
  $^1$Astrophysics Group, Keele University, Keele, 
      Staffordshire ST5 5BG\\
$^2$MIT Kavli Institute for Astrophysics and Space Research, Massachusetts Institute of Technology, Cambridge, MA 02139, USA
}

\date{13th September 2023}

\pagerange{\pageref{firstpage}--\pageref{lastpage}} \pubyear{2023}

\def\LaTeX{L\kern-.36em\raise.3ex\hbox{a}\kern-.15em
    T\kern-.1667em\lower.7ex\hbox{E}\kern-.125emX}

\begin{document}
\label{firstpage}
\maketitle

\begin{abstract}
IC~4665 is one of only a dozen young open clusters with a ``lithium depletion boundary" (LDB) age. Using an astrometrically and spectroscopically filtered sample of cluster members, we show that both the positions of its low mass stars in {\it Gaia} absolute colour-magnitude diagrams and the lithium depletion seen among its K- and early M-stars are discordant with the reported LDB age of $32^{+4}_{-5}$\, Myr. Re-analysis of archival spectra suggests that the LDB of IC~4665 has not been detected and that the published LDB age should be interpreted as a lower limit.  Empirical comparisons with similar datasets from other young clusters with better-established LDB ages indicate that IC~4665 is bracketed in age by the clusters IC~2602 and IC~2391 at $55 \pm 3$\, Myr.

\end{abstract}


\section{Introduction}

Determining the ages of young, low-mass stars is an important but difficult task. Young clusters and associations play the key role in calibrating both stellar evolutionary models and a variety of age-sensitive empirical diagnostics that can be used to estimate the ages of other stars \citep[e.g., see][]{Hillenbrand2009a, Soderblom2010a, Soderblom2014a, Jeffries2014b, Barrado2016b}. However, the {\it absolute} age scale for pre main-sequence (PMS) stars is still uncertain by factors of two or more \citep[e.g.,][]{Naylor2009a, Bell2013a, Mamajek2014a, Feiden2016a, Fang2017a, Jeffries2017a}, with significant disagreements between ages derived using different models and methods or even from stars in the same cluster with different masses. This in turn means there are still significant uncertainties in the timescales of all the evolutionary processes involved in star and planet formation -- for example the timescales for the dispersal of circumstellar material -- since these are determined using PMS stars in clusters and associations with estimated ages \citep[e.g.,][]{Haisch2001a, Fedele2010, Richert2018a}.

Of the age determination techniques used in young clusters, the "Lithium Depletion Boundary" (LDB) method \citep[][]{Bildsten1997a, Barrado1999a, Stauffer1999a, Jeffries2005a} -- finding the luminosity at which fully convective contracting PMS stars achieve the core temperatures necessary to rapidly burn their initial Li -- is likely to be the least model-dependent. LDB ages are less sensitive to changes in adopted nuclear reaction rates and opacities, or to the treatment of convection, rotation, magnetic fields and starspots than are ages from the Hertzsprung-Russell diagram \citep[][]{Burke2004a, Soderblom2014a, Jackson2014b, Tognelli2015a}.

Measuring LDB ages is difficult, involving establishing the presence or not of lithium in relatively high resolution spectra of very faint, low-mass M-dwarfs. Only a dozen or so clusters and moving groups have LDB ages and these could form the basis of a robust absolute age scale between about 10-700 Myr \citep[see][and references therein]{Randich2018a, Martin2018a, Binks2021a, Galindo2022a}. One of these clusters is IC~4665 ($=$C1746+056), for which \cite{Manzi2008a} found an LDB age of $28 \pm 5$\,Myr. Re-determinations of the LDB age, using the same reported Li measurements of Manzi et al., but a range of bolometric corrections and evolutionary models, arrive at ages of 20-32\,Myr \citep[][]{Soderblom2014a, Randich2018a, Galindo2022a}.
IC~4665 appears to be one of few clusters with an LDB age $<40$ Myr
and it has become an exemplar and calibrating object for stars of this age \citep[][]{Cargile2010b, Smith2011a, Meng2017a, Randich2018a, Miret-Roig2020a}. 

There is however some controversy over the age of IC~4665. There is more Li depletion among its K- and early M-type stars than expected when compared to other clusters with LDB ages of 20--40\,Myr, suggesting it could even be as old as 100 Myr \citep{Martin1997a, Jeffries2009a}; \cite{Jeffries2023a} fitted an empirical Li-depletion model to similar data from the {\it Gaia}-ESO survey, obtaining an age of  $52^{+8}_{-6}$ Myr.  This would contradict the LDB age determined by \cite{Manzi2008a} but would be consistent with the isochronal age of  $53^{+11}_{-9}$ Myr found in the large, homogeneously-fitted cluster sample of \cite{Dias2021a}. \cite{Cargile2010a} also noted that their derived high-mass stellar turn-off age of $42 \pm 11$ Myr, whilst formally consistent with the LDB age, did not follow the pattern in other young clusters that LDB ages were found to be {\it older} than turn-off ages \citep[e.g.,][]{Stauffer1998a, Stauffer1999a}.

In this paper we revisit the age of IC~4665 using: a revised distance and membership determination and absolute colour-magnitude diagrams from {\it Gaia} Data Release 3 \citep[{\it Gaia}~DR3,][]{Gaia2016b, Gaia2023a}; new membership and lithium data from the {\it Gaia}-ESO survey \citep[GES,][]{Gilmore2022a, Randich2022a}; and a re-examination of the spectroscopy used by \cite{Manzi2008a}. We conclude that IC~4665 has an age of 50--60 Myr and that its published LDB age is a lower limit.

\section{Cluster membership, distance and extinction}

\subsection{IC4665}

\label{IC4665mem}

\begin{figure}
    \centering
    \includegraphics[width=0.46\textwidth]{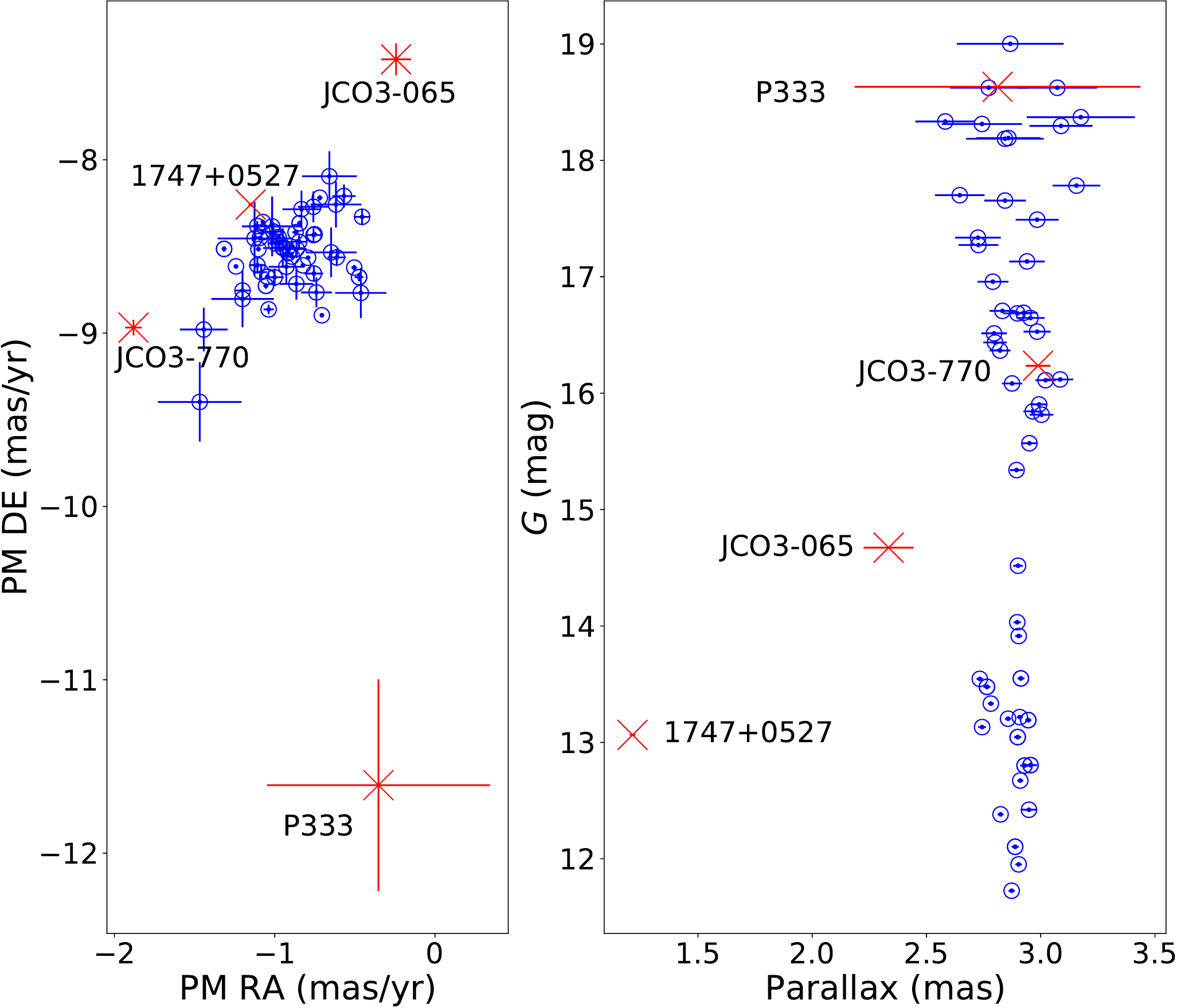}
\caption{IC 4665 candidates in the proper motion plane (left) and a plot of parallax versus apparent $G$ magnitude (right). Stars that were discarded as outliers in either plot are shown as red crosses and named according to the catalogues of \protect\cite{Jackson2022a} and \protect\cite{Galindo2022a} (see \S\ref{IC4665mem}).}
\label{Fig_pmplx}
\end{figure}

\begin{table}
\caption{The basic data for the stars assigned cluster membership in IC~4665 and each of the comparison clusters. The format and content of the Table is given here; the full table, consisting of 537 rows, is available in electronic format.}
\begin{tabular}{|l|l|l|}
\hline
Col  & Label & (Units)/Description \\
\hline
1 & Cluster & Cluster identifier \\
2 & Target  & Name \citep{Jackson2022a, Galindo2022a} \\
3 & RA & (degrees) J2000.0 \\
4 & DEC & (degrees) J2000.0 \\
5 & Mem & Membership probability \citep{Jackson2022a} \\ 
6 & Teff & (K) Effective temperature\\
7 & l\_EWLi & Upper limit flag on EWLi \citep[][]{Galindo2022a} \\ 
8 & EWLi & (m\AA) \\
9 & EWLic & (m\AA) corrected for blending \\
10 & e\_EWLic & (m\AA) \\
11 & DR3Name & Identifier in {\it Gaia} DR3 \\
12 & plx & (mas) parallax (from {\it Gaia} DR3) \\
13 & e\_plx & (mas) uncertainty in parallax \\
14 & dist & (pc) inferred distance \\
15 & e\_dist & (pc) \\
16 & Gmag & (mag) $G$ magnitude (from {\it Gaia} DR3) \\
17 & e\_Gmag & (mag) \\
18 & BPmag & (mag) $B_p$ magnitude (from {\it Gaia} DR3) \\
19 & e\_BPmag & (mag) \\
20 & RPmag & (mag) $R_p$ magnitude (from {\it Gaia} DR3) \\
21 & e\_RPmag & (mag) \\
22 & phot\_xs & Ratio of corrected flux excess factor to its dispersion \\
23 & Gmag0 & (mag) Extinction-corrected $G$ magnitude \\
24 & MG & (mag) Absolute $G$ magnitude \\
25 & e\_MG & (mag) \\
26 & BPRP0 & (mag) intrinsic $B_p-R_p$ colour \\
27 & e\_BPRP0 & (mag) \\
28 & GRP0 & (mag) intrinsic $G-R_p$ colour \\
29 & e\_GRP0 & (mag) \\
\hline
\end{tabular}
\label{tab_members}
\end{table}

For these investigations it is important to choose a secure sample of IC~4665 members that are not selected solely on the basis of their colours and magnitudes or because of their lithium abundance. To that end we compiled a list of members that have both spectroscopic and astrometric properties that make them very likely cluster members from \cite{Jackson2022a} and \cite{Galindo2022a}. The former were selected as GES targets with membership confirmed using {\it Gaia} DR3 proper motions along with GES radial velocities; IC~4665 targets with membership probability $>90$ per cent were included. The latter were taken from the previous spectroscopic surveys of \cite{Manzi2008a} and \cite{Jeffries2009a} but were screened by Galindo-Guil et al. using {\it Gaia} DR2 data to exclude objects with discrepant parallaxes and proper motions. 

Photometry from {\it Gaia} DR3 was added for all these members and equivalent widths of the Li~{\sc i}~6708\AA\ resonance line (EWLi) from (in preference order) \cite{Jeffries2023a}, \cite{Jeffries2009a} and \cite{Manzi2008a}. For seven stars observed in GES by both the medium resolution Giraffe spectrograph and at higher resolution with UVES \cite[see][]{Gilmore2022a}, a weighted mean EWLi was taken.
Effective temperatures ($T_{\rm eff}$) were taken from homogenised GES spectroscopic estimates \citep{Hourihane2023a} or the spectral energy distribution modelling of \cite{Galindo2022a}. Figure~\ref{Fig_pmplx} shows the proper motions and parallaxes for these stars. There are four clear outliers; three of them have a renormalised unit weight error RUWE$>1.4$ \citep[see][]{Lindegren2021a}, which may mean their astrometry is affected by 
binarity but they are removed at this stage. The outliers are: JCO3-770, JCO3-065 and P333, which were assigned membership by \cite{Galindo2022a} (though P333 lacked astrometry in {\it Gaia} DR2) and 17473947+0527101 from \cite{Jackson2022a}, where parallax was not used explicitly as a filter.  JCO3-065 has $T_{\rm eff} = 4460$\,K, EWLi $=91 \pm 22$\, m\AA\ \citep{Jeffries2009a} and is a candidate binary cluster member. The star P373 was also removed since it lacks a proper motion and parallax. The identifiers used in the membership papers, positions, EWLi, $T_{\rm eff}$ Gaia identifiers and photometry for the remaining 53 stars are listed in Table~\ref{tab_members} (available in electronic format only).

To plot an absolute colour magnitude diagram, the Gaia photometry was corrected for extinction. Estimates for IC~4665 include: a mean $E(B-V)=0.18 \pm 0.01$ from $uvby\beta$ photometry, \citep[with any scatter limited to $<0.04$ mag,][]{Crawford1972a}; $E(B-V) = 0.22^{+0.05}_{-0.06}$ from fitting isochrones to {\it Gaia} DR1 data \citep{Randich2018a}; $E(B-V)=0.23 \pm 0.03$ from {\it Gaia} DR2 photometry and isochrone fitting \citep{Dias2021a}; and $E(B-V) = 0.15 \pm 0.02$ from comparing {\it Gaia} DR3 colours with model isochrones \citep{Jackson2022a}. We adopt $E(B-V) = 0.19 \pm 0.04$ as a central value. Extinction in the {\it Gaia} DR3 bands was calculated and removed according to the methods described in \cite{Danielski2018a}, using the \cite{Fitzpatrick2019a} extinction law and updated for the DR3 photometry\footnote{\protect\href{https://www.cosmos.esa.int/web/gaia/edr3-extinction-law}{https://www.cosmos.esa.int/web/gaia/edr3-extinction-law}}. Photometry was flagged as poor quality if the ratio of the absolute value of the ``corrected flux excess factor" to its expected dispersion  was $>3$, and the $B_p-R_p$ colour was not used if $B_p > 20.3$ \citep[see][for details]{Riello2021a}.

Since the parallax uncertainties are generally small compared with the intrinsic dispersion within the cluster (at least for stars with $G<18$), individual stellar distances were estimated. This was done with a Monte-Carlo simulation of the distance likelihood distribution multiplied by a Gaussian prior probability function iteratively determined from the weighted mean and dispersion of all cluster members. The parallaxes were corrected for a small (mean $38\, \mu$as) colour-, ecliptic latitude- and magnitude-dependent zero-point error according to the prescription of \cite{Lindegren2021b}. Absolute magnitudes and intrinsic colours, corrected for extinction and individual distances, are given in Table~\ref{tab_members}. The weighted mean distance of the cluster members is $343.6 \pm 1.5$ pc.

\subsection{Comparison clusters}

\begin{table*}
\caption{The lithium depletion boundary age, reddening and derived properties of IC~4665 and the comparison clusters.}
\begin{tabular}{lccccccccc}
\hline
Cluster & LDB age$^1$ & $M_K$ (LDB)$^2$  & $E(B-V)$ & $N_{\rm mem}\,^{3}$ & Distance$^4$ & \multicolumn{2}{c}{Offsets (mag)$^5$} & \multicolumn{2}{c}{{\sc eagles} Li age$^6$} \\
        & Myr     & mag             & mag      &               & pc       &
$(B_p-R_p)/M_G$ & $(G-R_p)/M_G$ & \multicolumn{2}{c}{Myr} \\
\hline
IC~4665 & $31.6^{+3.9}_{-4.7}$ & $6.22\pm 0.13$ & 0.190 & 53 & $343.6 \pm 1.5$ & $-0.19\pm 0.03$ & $-0.13 \pm 0.03$ & $62.9^{+7.7}_{-6.2}$ & 52.9$^{+6.8}_{-5.2}$ \\
\\
NGC~2232 & $37.1^{+1.9}_{-1.9}$ & $6.89 \pm 0.10$ & 0.070 & 71 & $317.4 \pm 1.1$ & $+0.12 \pm 0.01$ & $+0.10\pm 0.02$ & $28.9^{+1.2}_{-1.0}$ & $29.3^{+1.2}_{-1.0}$ \\
NGC~2547 & $43.5^{+1.7}_{-4.0}$ & $6.94 \pm 0.12$ & 0.040 & 185 & $381.8 \pm 0.4$ & $+0.01 \pm 0.01$ & $-0.02 \pm 0.01$ & $34.6^{+1.2}_{-1.0}$ & $39.2^{+1.6}_{-1.3}$ \\
IC~2602  & $52.5^{+2.2}_{-3.7}$ & $7.35 \pm 0.11$ & 0.031 & 56 & $150.3 \pm 0.5$ & $-0.07 \pm 0.02$ & $-0.10 \pm 0.02$ & $45.8^{+4.7}_{-3.3}$ & $45.8^{+4.7}_{-3.3}$\\
IC~2391  & $57.7^{+0.5}_{-1.0}$ & $7.60 \pm 0.10$ & 0.030 & 48 & $150.0 \pm 0.4$ & $-0.15 \pm 0.02$ & $-0.21 \pm 0.03$ & $60.1^{+26.8}_{-9.2}$ &$60.1^{+26.8}_{-9.2}$ \\
Blanco~1 & $137.1^{+7.0}_{-33}$ & $8.88 \pm 0.16$ & 0.010 & 124 & $234.8 \pm 0.4$ & $-0.76 \pm 0.02$ & $-0.79 \pm 0.01$ & $68.2^{+9.2}_{-5.6}$ & $73.5^{+12.9}_{-6.9}$ \\
\hline
\multicolumn{10}{l}{$^1$ Using the BT-Settl or \protect\cite{Baraffe2015a} (for NGC~2232) models; taken from \protect\cite{Binks2021a} and \protect\cite{Galindo2022a}.}  \\
\multicolumn{10}{l}{$^2$ Using literature values for $K_{\rm LDB}$, corrected for extinction and the distance derived here.}\\
\multicolumn{10}{l}{$^3$ Total number of members after the filtering described in \S\ref{IC4665mem}}.\\
\multicolumn{10}{l}{$^4$ Weighted mean distance of the cluster members considered here.}\\
\multicolumn{10}{l}{$^5$ Offsets from the fiducial NGC~2547 isochrone in \S\ref{secABSCMD}.}\\
\multicolumn{10}{l}{$^6$ The second number shows the most probable age with $>3$-sigma outliers removed.}
\end{tabular}
\label{tab_clusters}
\end{table*}

\begin{figure}
    \centering
    \includegraphics[width=0.45\textwidth]{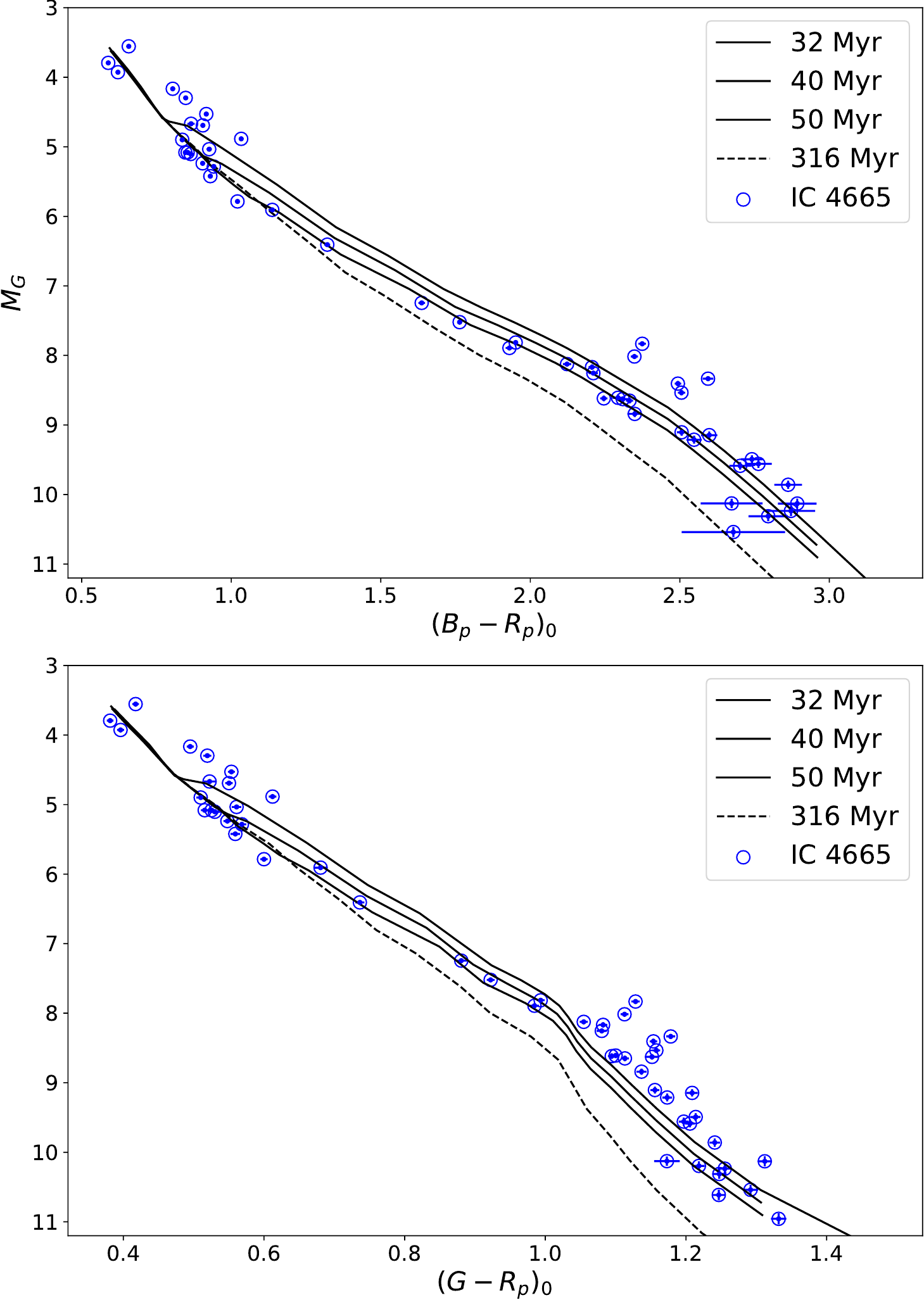}
\caption{Absolute colour magnitude diagrams for members of IC~4665 in the Gaia DR3 photometric system. Also shown are model isochrones from \protect\cite{Somers2020a} for ages around that of IC~4665 and at an older age - approximating the zero age main sequence.}
\label{FigCMD}
\end{figure}  

Since estimating young stellar ages from absolute colour magnitude diagrams or from lithium depletion in G-, K- and M-type stars is highly model dependent \citep[see][]{Soderblom2014a}, we compiled comparison samples from several other young clusters with well-established LDB ages, that were also observed in GES and which have homogeneous membership determination \citep{Jackson2022a} and EWLi measurements \citep{Jeffries2023a}. These clusters are NGC~2547, IC~2602, IC~2391, Blanco~1 and NGC~2232. The first four clusters have LDB ages of 44, 53, 58 and 137 Myr respectively, homogeneously determined by \cite{Galindo2022a}, based on spectroscopy originally presented by \cite{Jeffries2005a}, \cite{Dobbie2010a}, \cite{Barrado2004a} and \cite{Cargile2010a}, and using the BT-Settl models \citep{Allard2012b}. The same authors and models give an age of 32 Myr for IC~4665. Reddening for these clusters was also taken from \cite{Galindo2022a}. The LDB age of NGC~2232 was determined by \cite{Binks2021a} as $37 \pm 2$ Myr using \cite{Baraffe2015a} models, and we assume a reddening $E(B-V)=0.07 \pm 0.02$ from \cite{Lyra2006a}. Relevant literature properties and those subsequently derived here for these clusters and IC~4665 are listed in Table~\ref{tab_clusters}.

Membership lists for the comparison clusters were compiled and filtered as in \S\ref{IC4665mem}, combining the GES data with the tables in \cite{Galindo2022a}. Only GES membership and EWLi are available for NGC~2232. Screening on {\it Gaia} DR3 proper motions and parallaxes resulted in the removal of 7 , 6, 1, 6 and 3 targets from NGC~2547, IC~2602, IC~2391, Blanco~1 and NGC~2232 respectively. Individual distance estimations for the members, correction for extinction and individual distances was also done as described in \S\ref{IC4665mem}. The relevant data for individual cluster members are included in Table~\ref{tab_members} and the weighted mean cluster distances are in Table~\ref{tab_clusters}.

\section{Absolute Colour-Magnitude diagrams}

\label{secABSCMD}

\begin{figure*}

    \includegraphics[width=0.46\textwidth]{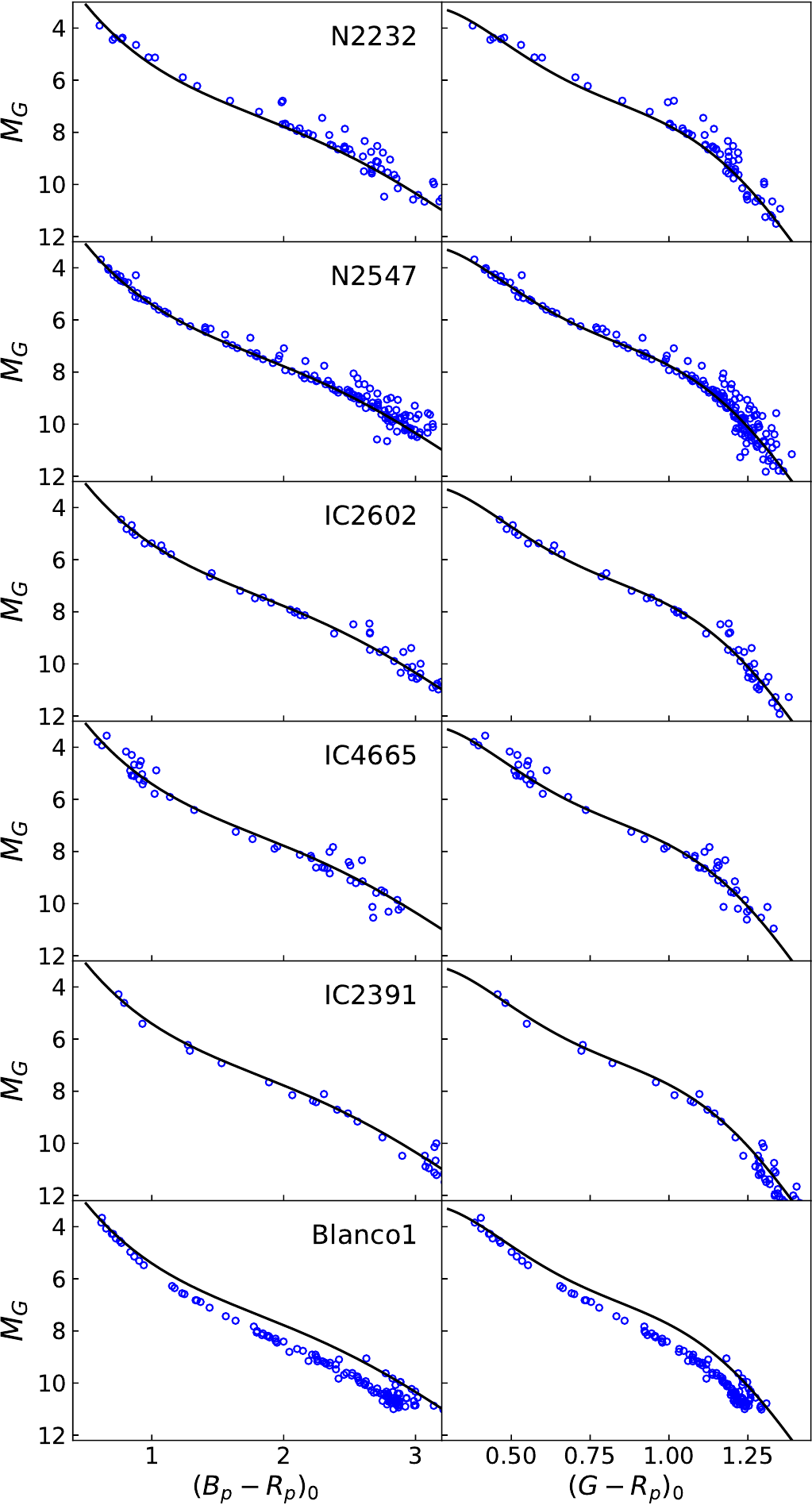}
    \hspace{5mm}
    \includegraphics[width=0.466\textwidth]{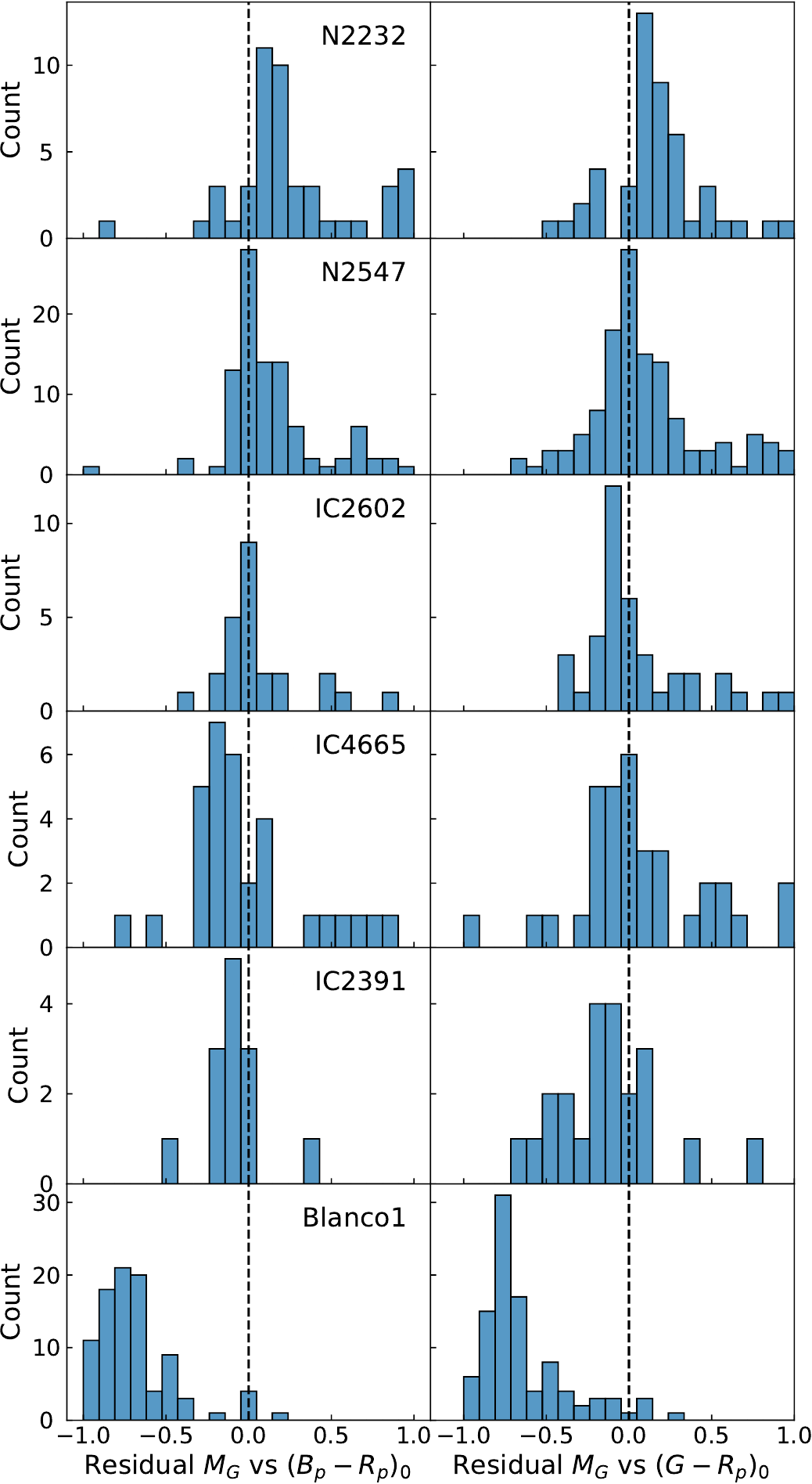}
\label{Fig_allCMDs}
\caption{(Left) The absolute colour magnitude diagrams of IC~4665 and comparison clusters. In each plot the solid line shows an empirical, fiducial isochrone fitted to the single-star sequence of NGC~2547. The clusters are presented in what appears from the CMDs to be their age order (youngest at the top). (Right) The corresponding histograms showing the distribution of residuals to the fiducial isochrone in each CMD. A negative residual means the star is fainter (older) than the fiducial isochrone at its colour.}
\end{figure*}

The absolute colour-magnitude diagrams (CMDs) for IC~4665 are shown in Fig.~\ref{FigCMD} along with several theoretical isochrones taken from the {\sc spots} models of \cite{Somers2020a}. The unspotted versions (i.e. stars without starspots) do a reasonable job of matching the IC~4665 single star sequence in the $(B_p-R_p)_0/M_G$ diagram at an age of $\sim 50$ Myr but do not match the shape of the observed sequence in $(G-R_p)_0/M_G$. The purpose of this comparison is not to estimate directly an isochronal age but to show which colour ranges are expected to be age sensitive ($(B_p-R_p)_0 > 0.98$ and $(G-R_p)_0 > 0.57$) and show that the isochrones are close-to-parallel in the colour-ranges that are age-sensitive, with a mean separation in $M_G$ from 30-50\,Myr of $\sim 0.015$ mag/Myr.

Different models would yield different age estimates and with widely varying success at matching the cluster sequence across the mass-range \citep[e.g.,][]{Binks2022a}. However, the separation between isochrones at a given age is much less model-dependent. Our approach therefore is to design empirical, fiducial isochrones in the CMDs that match the observed data and then to see by how far IC~4665 and the comparison clusters are offset from this isochrone as a means of establishing their age sequence and of estimating more mildly model-dependent age differences. 

Since it has $\sim 2.5$ times as many members as IC~4665 and a much lower reddening (and reddening uncertainty), we use NGC~2547 to establish the fiducial isochrones in both CMDs. We remove one third of the cluster members that lie above the clear single-star sequence as potential binary systems and then fit the remaining stars with fourth and fifth order polynomials in the $(Bp-Rp)_0/M_G$ and $(G-R_p)_0/M_G$ CMDs respectively. The results are shown in Fig.~\ref{Fig_allCMDs}. Note that only stars with ``good" photometry (as defined in \S\ref{IC4665mem}) were used in the fit and plotted.

The absolute CMD of IC~4665 is also shown in Fig~\ref{Fig_allCMDs} with {\it the same} fiducial isochrone that fits the NGC~2547 single star-sequence. A clear impression is gained that the IC~4665 single-star sequence lies {\it below} that of NGC~2547 in both CMDs and therefore that IC~4665 is {\it older} than NGC~2547. To quantify this, the right-hand panels of Fig.~\ref{Fig_allCMDs} show histograms of the residuals in absolute magnitude to the fiducial isochrones in both CMDs. Only stars in common, age-sensitive ranges of $0.98< (B_p-R_p)_0 < 2.9$ and $0.57<(G-R_p)_0<1.3$ are included. If we discard the upper third of the residuals as potential binaries and calculate the weighted means of the remainder (with error bars that account for uncertainties in both colour and absolute magnitude) we obtain $\Delta M_G = +0.01 \pm 0.01$ mag and $-0.02 \pm 0.01$ mag for NGC~2547 in the $(B_p-R_p)_0/M_G$ and $(G-R_p)_0/M_G$ CMDs respectively, where a positive residual is brighter/younger than the fiducial isochrone. These are close to zero, as expected. However, for IC~4665 the equivalent weighted mean residuals are $-0.19 \pm 0.03$ mag and $-0.13 \pm 0.03$ mag respectively, indicating that IC~2547 may be $\sim 10$ Myr older than NGC~2547. The error bars here are merely statistical and we also considered how uncertainties in reddening affect the result. This uncertainty is largest for IC~4665 at $\pm 0.04$ in $E(B-V)$. Perturbing the assumed reddening by this amount and repeating the exercise, we find that the $\Delta M_G$ values for IC~4665 change by only $\mp 0.03$ mag. Thus the result appears robust to statistical uncertainties and, given the homogeneity of the photometry, astrometry and membership selection, robust to systematic uncertainties too. 

This experiment was repeated for the other comparison clusters and the weighted mean offsets from the fiducial isochrone and statistical uncertainties are summarised in Table~\ref{tab_clusters}. The absolute CMDs and histograms of residuals are shown in Fig.~\ref{Fig_allCMDs}. The results show that the age order of the comparsion clusters (from youngest to oldest) is NGC~2232, NGC~2547, IC~2602, IC~2391 and Blanco~1, with IC~4665 falling between IC~2602 and IC~2391.

\section{Lithium depletion in the G-, K- and M-stars of IC~4665}

\label{secLi}

\begin{figure*}
    \centering
    \includegraphics[width=\textwidth]{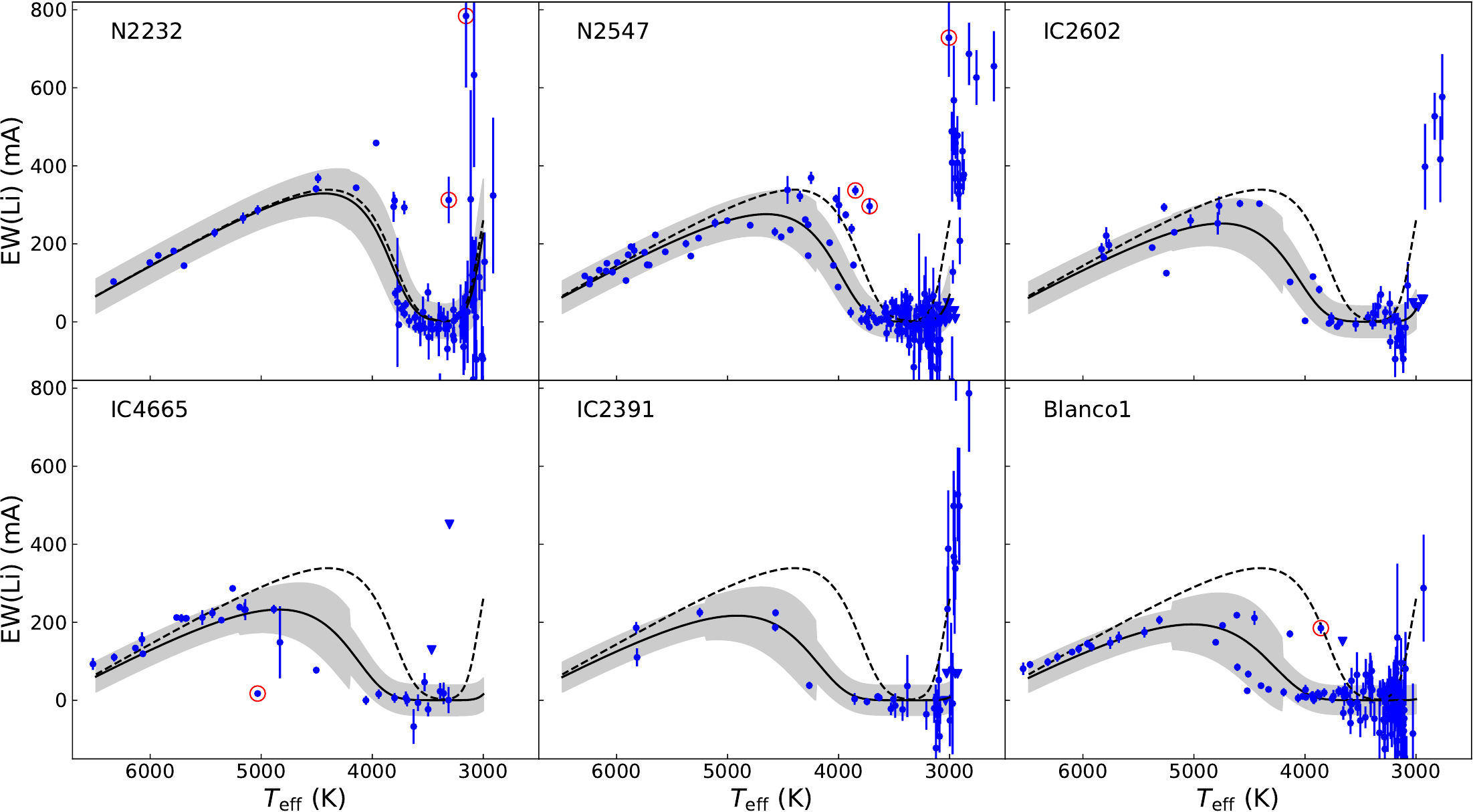}
\label{Fig_Li}
\caption{The Li depletion patterns of IC~4665 and the comparison clusters. The plots show EW(Li) versus $T_{\rm eff}$ with triangles indicating upper limits. Points surrounded by red circles are those that were removed by 3-sigma clipping (see~\S\ref{secLi}). The solid lines and shaded areas represent the {\sc eagles} isochrone corresponding to the most probable age (the clipped values in Table~\ref{tab_clusters}) and the modelled intrinsic dispersion for each cluster. The dashed line is the same in each plot and is an {\sc eagles} isochrone at 28\,Myr -- the original published LDB age of IC~4665 \protect\citep{Manzi2008a}.
}
\end{figure*}

A similar exercise was undertaken using Li-depletion among the G- to M-type stars in each of the clusters. In this case, a set of empirical isochrones in the EWLi versus $T_{\rm eff}$ plane has already been calibrated using all the data for open clusters observed as part of GES and implemented in the {\sc eagles} code \citep{Jeffries2023a}. This code models a cluster of stars by fitting the empirical isochrones, and a modelled intrinsic dispersion around those isochrones, for EWLi-$T_{\rm eff}$ data in the range $3000< T_{\rm eff}/{\rm K} < 6500$. A combined likelihood for all the stars is multiplied by a prior probability that is flat in $\log$ age to form a posterior probability distribution of age.

In IC~4665 and each of the comparison clusters, {\sc eagles} was used to estimate the most probable age and an asymmetric 68 per cent confidence interval from the selected members and these are reported in Table~\ref{tab_clusters}\footnote{The posterior probability distributions are all well constrained and similar to normal distributions in $\log$ age.}. Only stars with $3000 < T_{\rm eff}/{\rm K} < 6500$ are used in the fit. Where EWLi upper limits are quoted \citep[in the data retrieved from the ][catalogue]{Galindo2022a}, these were inserted at half the value of the upper limit, with a 1-sigma error bar equal to half the upper limit. For the few stars where the EWLi measurements originate from sources cited in \cite{Galindo2022a}, these were corrected for the unresolved blend with a close Fe~{\sc i} line using the formula given in \cite{Jeffries2023a}. In IC~4665, the very cool stars studied by \cite{Manzi2008a} have no quoted EWLi, only an indication of whether Li was detected or not. These were not included and are discussed in \S\ref{secLDB}. 

The ages and uncertainties determined in this way may not be robust to outliers - there is still some small possibility of contamination by non-members and the far wings of the intrinsic EWLi distribution around an isochrone are not well-constrained. We have added another column to Table~\ref{tab_clusters} showing the best-fit ages when 3-sigma outliers (including the observational error bar and the intrinsic dispersion) are clipped from the sample. It is these fits that are shown in Fig.~\ref{Fig_Li}. IC~4665 has one very low point removed by this process\footnote{17455308+0536293: This star is otherwise unremarkable in terms of its kinematics, parallax, colour and magnitude.} and its removal reduces the best fit age by 10 Myr because it lies in an age-sensitive part of the diagram. The other clusters are less affected - NGC~2232 has two points clipped, NGC~2547 three, Blanco~1 one, and IC~2602 and IC~2391 have no points clipped. Note that the very sharp rise in EWLi at $T_{\rm eff} \leq 3000$ K for ages $>30$ Myr, corresponding to the LDB, is not represented in the GES data used to train the {\sc eagles} model.

\section{Revisiting the LDB age of IC~4665}

\begin{figure*}
    \centering
    \includegraphics[width=\textwidth]{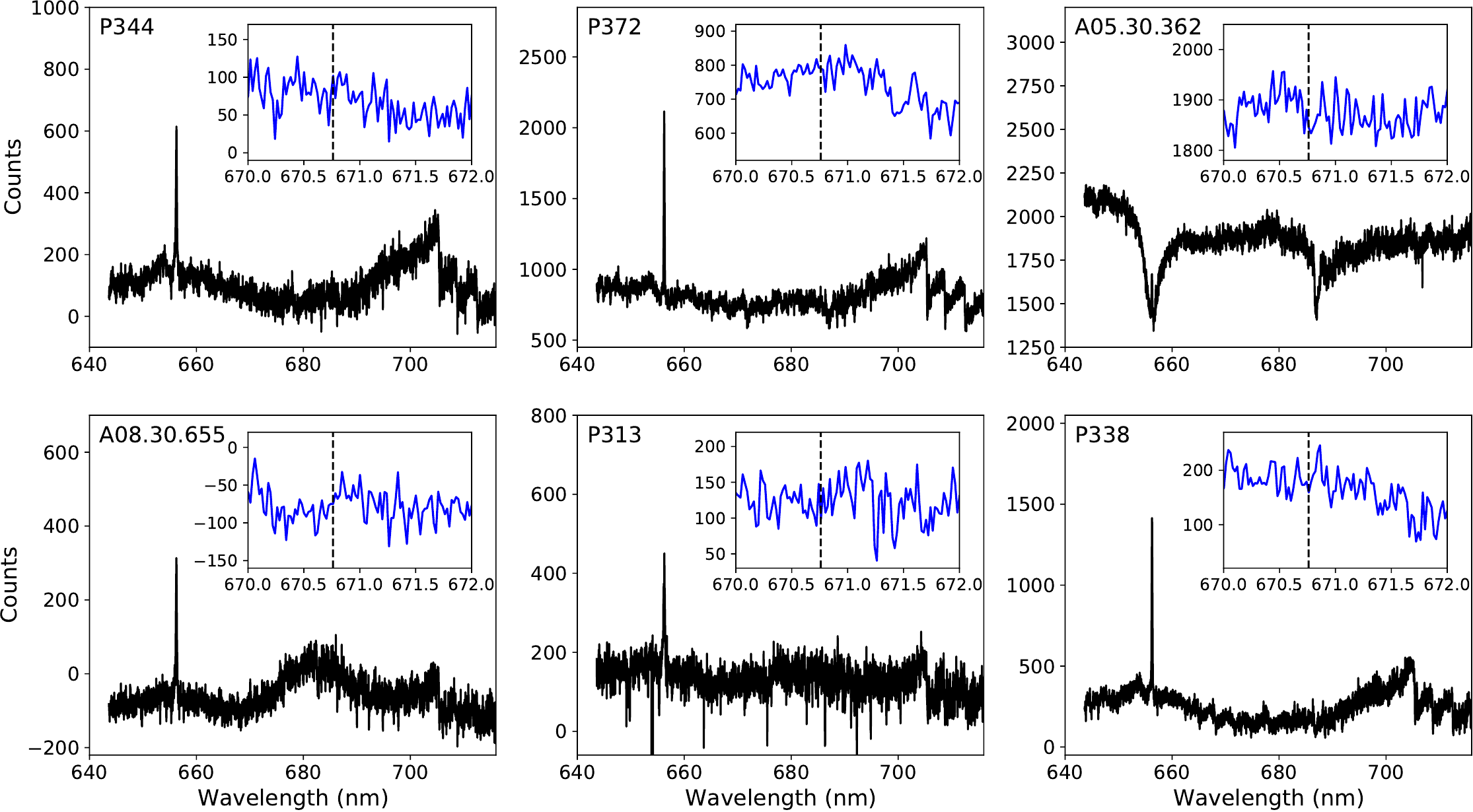}
\label{Fig_LDB}
\caption{Archival spectra of six low-mass IC~4665 members where a detection of the 670.8\,nm line was claimed in \protect\cite{Manzi2008a}. The pipeline-processed spectra have had the median sky spectrum subtracted (see \S\ref{secLDB}). The insets show the region around the 670.8\,nm Li{\sc i} feature.
}
\end{figure*}  

\label{secLDB}

\cite{Manzi2008a} reported an LDB age for IC~4665 based on detecting Li in five candidate low-luminosity members (P338, P333, P313, A.08.30.655, A.05.30.3622), the probable detection of two others at slightly higher luminosity (P372, P344) and then no Li detected in many other stars at even higher luminosities. The candidate members were confirmed with radial velocities and the presence of H$\alpha$ in emission. In their {\it Gaia}-DR2 astrometric study, \cite{Galindo2022a} kept all of the Li-rich low-luminosity sources except P333 for which no astrometry was available. In our {\it Gaia} DR3 filtering P333 is excluded on the grounds of proper motion (see Fig.~\ref{Fig_pmplx}) but the other claimed Li-rich sources are kept.

 The spectra shown in fig~2 of \cite{Manzi2008a} allow rough estimates of EWLi in the six Li-rich candidates, albeit the authors pointed out that sky subtraction problems in these fibre-based spectra may make the continuum levels unreliable. EWLi in the plotted spectra are $\sim 100-150$ m\AA, apart from P338 where EWLi $\sim 400$ m\AA. But there are numerous other features with EW$>300$ m\AA\ in the spectrum of P338 that are stronger or have no counterparts in similar spectra, and so we think this Li detection, and the others, could just be noise. In other clusters where stars below the LDB have been confirmed, the typical EWLi is $>300$ m\AA\ and usually $\geq 500$ m\AA\  \citep[e.g.,][and see Fig.~\ref{Fig_Li}]{Stauffer1998a, Barrado2004a, Jeffries2005a, Dobbie2010a}. Since EWLi in undepleted, very young mid-M dwarfs is empirically found to be $\sim 600$ m\AA\ \citep{Jeffries2023a} and total Li depletion is very rapid once it begins in fully convective stars \citep[a few Myr,][]{Bildsten1997a}, then a population of objects with low, but non-zero EWLi is not expected.

To investigate further, we recovered ``phase 3" pipeline-processed spectra of these six objects from the ESO archive. All were observed in one fibre setup using the Giraffe fibre spectrograph on the VLT-UT2 in August and September 2004 \citep[see][for details]{Manzi2008a}. There were six separate exposures, each amounting to about 40 minutes, but we discarded the first since it was badly affected by scattered light from an extremely bright star observed through one of the fibres. The remaining five exposures are also affected to a lesser extent by scattered light from a pair of targeted stars that are almost 8 magnitudes brighter than the LDB candidates. 

The pipeline-processed spectra are already debiased, flat-fielded and wavelength calibrated but not sky-subtracted. We attempted sky-subtraction by removing the median of the sky signal observed through 16 dedicated fibres placed on random blank sky regions. The results were mixed; whilst the numerous bright night-sky emission lines were successfully removed, indicating that the relative fibre transmission factors were well-calibrated, the continuum levels among different sky spectra varied by almost a factor of two and in some cases the sky-subtracted target flux was negative. The summed, sky-subtracted spectra are shown in Fig.~\ref{Fig_LDB}. For three targets (P344, P372, P338) it is clear from the TiO molecular absorption bands at $>705$\,nm that they are M-dwarfs and show (chromospheric) H$\alpha$ emission, as expected from young cool stars. Two other targets (A08.30.655 and P313) have very weak signals, but do exhibit H$\alpha$ emission. AO5.30.362 has an unusual spectrum that appears to be totally dominated by contaminating light from a very bright star whose spectrum is adjacent on the CCD detector and little further can be said about it.

The continuum levels of the sky-subtracted spectra are highly uncertain (and even negative) so reliable EWLi estimates are impossible. Nevertheless, we looked for evidence of any Li absorption feature by subtracting the flux in a 0.1 nm interval\footnote{This is roughly the FWHM expected for an unresolved line.} around 670.8 nm from the flux expected in a flat continuum defined by two windows ($\lambda\lambda$ 670.0-670.5\,nm and $\lambda\lambda$ 671.0-671.5\,nm) either side, and expressed as a multiple of the expected uncertainty based on the rms signal in the continuum windows.  The results varied from $+1.8$ for A05.30.362 (the positive indicating absorption) to $-1.5$ for P372. i.e. We have no evidence for any significant absorption feature with above 2-sigma confidence (see the insets in Fig.~\ref{Fig_LDB}). This is unlikely to change if there is a scattered light pedestal to be added/subtracted from the spectrum or if the subtracted sky were scaled in some way, since there is no significant absorption feature in the sky spectrum near the Li wavelength. For completeness, we confirmed this is also the case if the line integration window is broadened to 0.2\,nm, which might be more optimal for low-mass stars rotating at $\sim 60$ km\,s$^{-1}$.

If none of these faint stars has significant Li absorption, or even if it is unclear whether they have Li or not, then the LDB age of IC~4665 determined by \cite{Manzi2008a} and re-determined by \cite{Galindo2022a}, based on Manzi et al.'s identification of Li-rich stars, is a lower limit. The absolute $K$ magnitude of the LDB (see Table~\ref{tab_clusters}) is $M_K > 6.22$ and the LDB age would be $>32$~Myr.

\section{Discussion and Conclusions}

Prior to this paper, the LDB age analysis of \cite{Galindo2022a}, based on the Li detections reported by \cite{Manzi2008a}, found IC~4665 to be the youngest of the six clusters considered here (see Table~\ref{tab_clusters}). . 
This is confirmed in a model-independent way using the observed apparent $K_s$ magnitudes of the reported LDBs and correcting these for the redetermined {\it Gaia}-DR3 distances and reddening values listed in Table~\ref{tab_clusters}. The $M_K$ of the LDB inferred for IC~4665 is $>0.6$ mag brighter than any of the other clusters, a difference that cannot be explained by any plausible uncertainties in the cluster distances or extinctions. 

However, comparison of the absolute CMDs in \S\ref{secABSCMD} shows that the single star sequence in IC~4665 lies between those of IC~2602 and IC~2391 and $\sim 0.15$ mag fainter in $M_G$ than a fiducial isochrone that matches NGC~2547. This suggests that IC~4665 is $\sim 10$ Myr older than NGC~2547, which would be congruent with the difference between the LDB ages of NGC~2547 and those of IC~2602/2391. Similarly, the empirical ages determined from the {\sc eagles} model of Li depletion in the G-, K- and early M-stars of these clusters in \S\ref{secLi} also suggests that IC~4665 is at least 10~Myr older than NGC~2547 and bracketed by IC~2602 and IC~2391\footnote{Or is similar in age to IC~2391 if the 3-$\sigma$ clipped star is included.}.

This leads us to doubt whether \cite{Manzi2008a} successfully detected the LDB of IC~4665 at $M_K \sim 6.22$. In \S\ref{secLDB} we failed to find significant Li absorption in any of the archival spectra of the lowest luminosity IC~4665 members and this would indicate that the LDB age determined by \cite{Manzi2008a} and re-evaluated by \cite{Galindo2022a} should be interpreted as a {\it lower limit} of $>32$ Myr. We conclude that the age of IC~4665 is most likely between that of IC~2602 and IC~2391 at $55 \pm 3$ Myr, on the age scale defined by the homogeneous LDB determinations of \cite{Galindo2022a} (using the BT-Settl evolutionary models). We predict, based on the absolute magnitude of the LDB in the comparison clusters, that the LDB of IC~4665 will be found at $K_s \sim 15.2 \pm 0.2$, almost a magnitude fainter than the faintest stars investigated by \cite{Manzi2008a}.  Comparison with IC~2391, where the LDB was found at $R \sim 18.1$ \citep{Barrado2004a}, and taking account of the differences in distance and reddening, suggests that the LDB of IC~4665 will be found at $R \simeq 20.2$. This is a little brighter than the apparent magnitude of the faintest LDB observed so far ($R \simeq 20.8$) in the very young but distant cluster, NGC~1960 \citep{Jeffries2013a}. The latter LDB detection needed about 9 hours of spectroscopy on the 8-m aperture Gemini-North telescope; we anticipate that a successful LDB detection in IC~4665 may require similar resources.

An older age for IC~4665 is consistent with the levels of rotation and magnetic activity in its lower mass stars and with the turn-off age derived from its sparsely populated upper main sequence \citep[$42 \pm 11$ Myr,][]{Cargile2010b}. The older age should be considered when interpreting the fraction of stars possessing disks, either primordial or from collisional debris \citep{Smith2011a, Meng2017a, Miret-Roig2020a}. 
For instance, \cite{Meng2017a} and \cite{Miret-Roig2020a} find that IC~4665 has a lower disc fraction among its solar-type stars than NGC~2547 and other clusters with ages 10--30 Myr. However, this may not be a surprising result if IC~4665 is older than NGC~2547, since it is likely that the debris disc luminosities and hence observed disk frequencies decrease with age \citep{Siegler2007a, Pawellek2021a}.

\section*{Acknowledgments}
This work has made use of data from the European Space Agency (ESA) mission
{\it Gaia} (\url{https://www.cosmos.esa.int/gaia}), processed by the {\it Gaia}
Data Processing and Analysis Consortium (DPAC,
\url{https://www.cosmos.esa.int/web/gaia/dpac/consortium}). Funding for the DPAC
has been provided by national institutions, in particular the institutions
participating in the {\it Gaia} Multilateral Agreement.
Giraffe spectra were obtained from the ESO Science Archive Facility with DOI(s): \url{https://doi.eso.org/10.18727/archive27}.
This research has made use of the SIMBAD database, operated at Centre de Donn\'ees astronomiques de Strasbourg (CDS), France.

\section*{Data Availability Statement}
The data used in this paper are all publicly available from: the cited publications in the case of spectral parameters and cluster membership; from CDS: \url{https://cdsarc.cds.unistra.fr/viz-bin/cat/I/355} in the case of Gaia DR3 data; and  from the ESO phase 3 archive for the IC4665 Giraffe spectra: \url{https://archive.eso.org/wdb/wdb/adp/phase3_main/query?prog_id=073.D-0587}

\appendix

\bibliographystyle{mn2e.bst} 
\bibliography{references}

% \include{appendix.tex}

%%%%%%%%%%%%%%%%%%%%%%%%%%%%%%%%%%%%

\bsp % ``This paper has been produced using the ...''
\label{lastpage}
\end{document}